# Water under the Cover: Structures and Thermodynamics of Water Encapsulated by Graphene

Shuping Jiao[1], Chuanhua Duan[2] and Zhiping Xu[1,*]

[1]Applied Mechanics Laboratory, Department of Engineering Mechanics, and Center for Nano and Micro Mechanics, Tsinghua University, Beijing 100084, China

[2]Department of Mechanical Engineering, Boston University, Boston, MA 02215, USA

[*]Email: xuzp@tsinghua.edu.cn

## Abstract

Understanding the phase behaviors of nanoconfined water has driven notable research interests recently. In this work, we examine water encapsulated under a graphene cover, that widely appears in nanoelectronic devices based on two-dimensional materials and offers an ideal testbed to explore its molecular structures and thermodynamics. We find layered water structures up to ~1000 trapped water molecules, which is stabilized by the spatial confinement and pressure induced by the adhesion between graphene cover and substrate. For monolayer encapsulations, we identify two representative crystalline lattices for the two-dimensional ice, and the lattice defects. Free energy analysis shows that these structural orders with low entropy are compensated by high formation energies due to the pressurized confinement. There exists an order-disorder transition for this condensed phase at ~480-490 K, with a sharp reduction in the number of hydrogen bonds and increase in the entropy. Fast diffusion of the encapsulated water is identified, with an anomalous temperarture dependence that indicates the solid-fluid nature of this structural transition. These findings offer fundamental understandings of the encapsulated water, and provide guidance for practical applications with its presence, for example, in the design of two-dimensional materials based nanodevices with encapsulated water at their interfaces.

# I. INTRODUCTION

The formation mechanism of ordered and disordered hydrogen-bond (H-bond) networks has been a unsolved puzzle in understanding the structures and behaviors of water and ice [1]. Although significant attention has been paid to elucidate the phase diagram of bulk water in the temperature-pressure spaces, characterization of nanoconfined water at room temperature has only started recently. For example, water in chain and tubular forms are found inside carbon nanotubes [2]. Two-dimensional (2D) ordered H-Bond networks with various topologies are observed on solid surfaces [3] or within the nanoscale capillary galleries in layered materials [4]. The formation and stabilization of these ordered phases of water were proposed to benefit from the effects of nanoconfinement, lattice matching, surface interaction, and pressure [5-7]. This much enriched phase diagram of water in a confined space not only introduces new forms and thermodynamic behaviors of water, such as the 2D ice and fast mass transport in hydrophobic nanochannels [4], but also poses critical questions on the fundamental understanding of nanoconfined water.

As a finite system, the surface of small water droplets or clusters plays an important role in defining their thermodynamics [8]. Similarly, under nanoconfinement, the interface between water and structures not only modulates the H-bond network near the interface, leading to structural ordering such as layering or crystalization, but also applies an anisotropic pressure onto the water condensation. For example, in-plane pressure on the order of 1 GPa was estimated for water capillary confined between graphene oxide layers, although the out-of-plane can be absent in a relaxed structure [4]. In the development of nanoelectronic devices with thin films deposited onto substrates, intercalated water has been characterized at the interface [9-12]. Encapsulation by the wrinkles or ripples in the graphene membrane thus provides an isolated, strong nanoconfinement for water, which could then be used as a test chamber to probe the thermodynamics of water in this specific condition. Moreover, the encapsulated confinement is different from those in nanoslits, nanopores or on surfaces because of the presence of pressure in EWs resulted from the adhesion between graphene and the substrate, which thus adds new understandings to those on the existing models of nanoconfined water.

In this work, we explore the structures and thermodynamics of water encapsulated by a graphene layer deposited on a solid surface by performing molecular dynamics (MD) simulations. We first report the ordered and disordered molecular structures of water condensations in the encapsulation, and then explore their structural transitions and thermodynamics based on free energy analysis. We also analyze molecular and collective diffusion of EWs that demonstrates a strong correlation with their molecular structures.

## II. MODELS AND METHODS

*Molecular Structures*. Both 3D models with the number of water molecule $N_W$ = 80-1536 and 2D models with $N_W$ = 60-708 are constructed in this work. In the 3D model, the lateral size of the graphene sheets is 20x20 nm, and an open boundary is used in the $z$ direction. In the 2D model, the size of graphene sheet is 25x2.13 nm, and a periodic boundary condition (PBC) is applied only in the $y$ direction.

*Molecule Dynamics Simulations Details*. We perform molecular dynamics simulations using the large-scale atomic/molecular massively parallel simulator (LAMMPS) [13]. The all-atom optimized potentials for liquid simulations (OPLS-AA) are used for graphene [14-16]. TIP4P and SPC/E models of water are used for a comparative study, which were both widely adopted for MD simulations of water and its phase transition behaviors [17-19]. Our MD simulation results show that these two models predict the same water structure, but slightly different structural transition temperature. The interaction between graphene and water is described with the set of simulation parameters, $\varepsilon_{C-O}$ = 4.063 meV, $\sigma_{C-O}$ = 0.319 nm, which predicts a water contact angle (WCA) of $\theta_{c,G}$ = 98.4° for graphene that is in consistency with experimental measurements [16]. The long-range Coulomb interactions are computed by using the particle-particle particle-mesh algorithm (PPPM) [20]. The time step for the equation-of-motion integration is 1 fs., with the SHAKE algorithm applied for the stretching terms between oxygen and hydrogen atoms of water to reduce high-frequency vibrations that require a very short time step. Both the 3D and 2D structures are equilibrated using a Berendsen thermostat at 200-600 K for 2 ns. MD simulations with temperature up to $T$ =

1600 K are explored to explore the thermal stability of EWs. For the thermodynamics analysis of entropy and free energy, we carry out equilibrium MD simulation in a NVT ensemble using the Nosé-hoover thermostat for another 0.2 ns, where the MD trajectories are written out every 4 fs for 50 ps [21]. *Structure Factor Analysis*. The in-plane structure factors of layered EWs is calculated as [18, 22]

$$S(\mathbf{q}) = \left\langle \frac{1}{N_W} \left[ \left( \sum_{i=1}^{N_W} \cos(\mathbf{q} \cdot \mathbf{r}_i) \right)^2 + \left( \sum_{i=1}^{N_W} \sin(\mathbf{q} \cdot \mathbf{r}_i) \right)^2 \right] \right\rangle$$

Here $\mathbf{r}_j$ is position of the $j$-th oxygen atom in the plane, $N_W$ is the number of water molecules, and $\mathbf{q}$ is wavevector. $<\ldots>$ denotes time average in thermal equilibrium.

*Characterization of Molecular and Collective Diffusion.* The diffusion coefficient $D$ is calculated from the molecular trajectories of water by using the Einstein's definition relating the correlation function of atomic positions $\mathbf{r}_i$, or the mean-square distance (MSD), to the diffusivity $D = \lim_{t->\infty} <|\mathbf{r}(t) - \mathbf{r}(0)|^2>/2d_it$. Here $d_i$ is the dimension of space, $t$ is the simulation time, and $<\ldots>$ is the ensemble average. The collective motion of EW is calculated from its center-of-mass trajectories. In our simulations with a time span of a few nanoseconds, the MSD $<|\mathbf{r}(t) - \mathbf{r}(0)|^2>$ is calculated based on the time-series of all oxygen atom position, with the 1500 time-averages starting from different time point in the series.

## III. Results

*Structures of Ordered and Disordered Water.* To model the nanoconfined water, we encapsulate a water droplet between two graphene sheets, where the bottom layer is fixed, representing a solid substrate. The top layer covers the droplet and is free to deform, to accommodate the structural changes in water. Both 2D and 3D models are constructed with hemispheric and half-cylindrical encapsulation, respectively (**Fig. 1a**). The atomic structures are equilibrated in our MD simulations at specific temperature. The simulation results show that at room temperature, ordered structures form in the encapsulation. Layered structures are distinct for mono-, bi- and tri-layers but not for larger droplet with the number of water molecules $N_W$ beyond ~1000 (**Fig. 1b**). For our 3D models, the transition from mono-layer to bi-layer structures occurs at $N_W = $ ~165-180, and the latter structure further changes into tri-layer EW with $N_W$ increases to ~750-986. The density profiles also characterize weakened layering order when the size of EW increases from mono-layer to the bi- and tri-layers, demonstrating the increasing fluidity (**Fig. 2a**).

The presence of interlayered H-bonds within the EW and reduced pressure weaken the layered order (**Fig. 2b**). This effect is similar to the observations for water confined between planar surfaces, where the layered order is reduced with increasing interlayer distance, although the non-flat confinement in our models results in less ordered structures [7]. Moreover, in the equilibrium at room temperature, molecules in the monolayer EW are able to vibrate in the H-bond network, but rotation and diffusion within the encapsulation is not activated, indicating the nature of solidity, while in either bi- or tri-layers both diffusion and rotation are observed.

In addition to the layering order, in-plane molecular structures with regular H-bond network are also identified. For the water monolayer encapsulated at room temperature, we find ordered region with almost-perfect 2D lattices, disordered region between them that include grain boundaries, as well as point defects (**Fig. 1c**). The H-bonds align with the plane of water layers in general. Each of the water molecules in these lattice bonds with neighboring molecules through four H-bonds, following the ice rule [1]. Structural factor analysis shows a distinct near-square symmetry for monolayer EW, which is absent in the bi- and tri-layers (**Fig. 2c**), and this conclusion holds for the 2D encapsulations as well (**Fig. S1**). Bond length and angle analysis of the H-bonding network show an average H-bond length of $l_{O-O}$ ~0.278 nm, and the bond angles span over a wide region around $\theta_{O-H-O}$ ~160°.

In the crystalline domains, we identify two types of 2D lattices, which are shown in **Fig. 1c**. Several types of quasi-2D ice structures, including rhombic structures, have been observed recently between hydrophobic walls with the interlayer distance $d < 0.7$ nm. The structures characterized are mostly the same or similar (with different out-of-plane or lateral structures, depending on the model used) as the type II structure identified in this work although the confinement is quite different [23, 24], while type I has not been reported yet. The change in nanoconfining conditions also leads to different phase diagrams of water, which depends on the water models used in the simulations as well. Considering the wide appearance of EWs in the graphene-substrate setup, our findings here have significant practical implications in the design of nanoelectronics. The two lattices reported here both have four water molecules in the unit cell. The difference exists in that for type I, there are four almost-identical rhombuses, and while the lattice type II contains two squares and two rhombuses. These two types of lattices are also indicated by three peaks at about 75$^{\circ}$, 90$^{\circ}$, 105$^{\circ}$ in the distribution of O-O-O angles in our structural analysis. It is interesting that these two types of lattice are geometrically compatible and can co-exist, as we observed in the simulation snapshots. The mixing nature of type I and II in our MD simulations and presence of defects in the ordered H-bond network originate from the non-flat spatial confinement and energy barriers that prohibit further perfection of the defective structures.

*Thermodynamics of Encapsulated Water.* These evidences of ordered water structures inside graphene encapsulation at room temperature are interesting not only because its formation mechanism offers hints in the understanding of water, but also provides an ideal platform to explore thermodynamics of condensed matter in a nanoconfined space. When a graphene sheet is deposited onto the substrate, the competition between graphene elasticity and the substrate interaction results in an optimal conformation of the EW [25]. Adhesion between two graphene layers further leads to a pressure inside the encapsulation. From our MD simulation results, we find that the pressure is on the order of $P = 0.1$-1 GPa, which decreases with the number of water molecules therein (**Fig. 2b**) [7]. Here the pressure within EW is averaged over the atomic pressure tensor of all water molecules, which includes both kinetic energy and virial contributions and is anisotropic in this situation [13]. Consequently, it contains water-water and carbon-water interactions but not the carbon-carbon interaction, and could be decomposed into in-plane and vertical components, respectively. The amplitude of in-plane pressure is usually about half of the normal pressure. Considering the ultrahigh in-plane tensile stiffness of graphene $k =$ ~340-690 N/m and low bending stiffness of $D =$ ~1 eV [26], this pressure will bend the graphene sheet instead of stretching it, adapting to the deformation of EW. Our structural analysis supports this conclusion by showing that the bond length in graphene is almost identical to that in the undeformed structure. From the phase diagram of bulk water, we find its density $\rho_0$ is ~1.1-1.2 g/cm$^3$ at pressure $P$ = ~0.3-1 GPa [1], which aligns with our simulation results, $\rho$ = ~1.07-1.25 g/cm$^3$, although the H-bond network in the EW is different from that in the bulk.

The phase stability of ordered water structures is assessed here by performing MD simulations within a wide range of temperature. We find that the EW is stable up to the condition where the graphene structure becomes unstable (**Fig. S2**). In this work, we limit our discussion within the temperature range of $T$ = 200-600 K for the interest of practical application in nanoelectronic devices. We find that, as $T$ increases, the pressure gently increases from 0.81 to1.0 GPa for a monolayer EW with 115 water molecules, and the molecular volume increases slightly from $0.83V_0$ to $0.9V_0$. Here $V_0$ is the molecular volume in bulk water (**Fig. 3a**). However, there is a distinct reduction of the number of H-bonds at 480-490 K, implying a pronounced structural change in the EW (**Fig. 3b** and **3c**).

To obtain more insights into the thermodynamics of EW, we calculate the entropy based on the two-phase thermodynamics (2PT) model developed by Lin *et al.* [21], as well as the free energy $F = E$ - $TS$ with additional correction terms for the zero point energy (ZPE) and heat capacity. The entropy shows a discontinuous jump near $T$ = 480-490 K, signaling a first-order phase transition (**Fig. 4a**), which agrees with the observed breakdown of ordered H-bond network (**Fig. 3b**). We also find that the EW has a much lower entropy ($S$ = 48.4, 54.4 and 54.1 J/mol.K for mono-, bi- and tri-layers) than that of the bulk water ($S_0$ = 60.3 J/mol.K), which corresponds to high -$TS$ values in the free energy that must be compensated by the reduction in the total energy, $E$, of EW (**Fig. 4b**). The low entropy value of monolayer EW is comparable, and slightly lower than that of the intercalated water in the slit between planar graphene sheets, while the total energy and free energy are much lower (**Table 1**). Moreover, as $T$ increases from 200 to 600 K, the change in $E$ is much smaller than that in the free energy $F$, which keeps decreasing (**Fig. 4b**). The distinct changes in structures of monolayer EWs can also be characterized by the structural factor $S(\mathbf{q})$ summarized from the MD simulations of 3D and 2D structure (**Fig. S2**).

## IV. DISCUSSION

From our simulation results, we conclude that the formation of ordered structures in the EW arise from the pressure induced by encapsulation and the nanoscale confinement where the lattice of graphene and its adhesion offers a quasi-epitaxial template for the EW. To elucidate the dominating mechanism in the formation of 2D 'ice' structures, a simulation of intercalated water between two rigid parallel graphene sheets is also carried out. We find that a single layer 2D 'ice' forms even in the absence of pressure, with an interlayer distance $d$ = 0.64 nm. However, the in-plane order is lower than that in the EWs, as indicated by the structural factor analysis (**Fig. S3**). We also notice that, the pressure $P$ = ~0.22-0.46 GPa in the 2D encapsulation is lower than that in the 3D encapsulation due to the weaker confinement, and as a result, the EW structure is less ordered in the 2D. We further tune the interaction strength between water molecules and graphene, which adjusts the pressure in the nanoconfined water. The simulation results (**Fig. S4**) show that the in-plane lattice structures vanish as the van der Waals interaction decreases, where the confinement and pressure become weaker and lower. From these facts, we conclude that the nanoconfinement is critical for the formation of layer structures, while the pressure controls the in-plane order of the room-temperature 2D ice. Moreover, our additional simulation results for water encapsulated between graphene and silica surface do not show ordered 2D lattice of water molecules due to the hydrophilic nature and irregular atomic structures of silica.

The phase transition at 480-490 K is likely to be a solid-fluid transition in 2D based on our structural analysis. To further characterize this transition, we explore the dynamical behaviors of water molecules in EW by computing the mean square distance (MSD) of molecular diffusion from trajectories of the molecules or collective diffusion from the center-of-mass motion of EWs (**Fig. 5a**). The graphene substrate is constrained in the MD simulations here, and externals force corresponding to the constraint preserves the conservation of total momentum while large-amplitude displacement is allowed for the water structures. The results clearly demonstrate the appearance of fluidity above the transition temperature. Dramatic increase of the MSDs occurs in the temperature range of ~480-490 K, which indicates a phase transition. More interestingly, we find that the EWs could diffuse collectively under the graphene coating, mostly in translational motion and gentle libration. From the data MSDs, we calculate the diffusion coefficient $D$ of the collective motion of EW (**Fig. 5c**), which is on the order of $10^{-5}$ cm$^2$/s at room temperature. However, the value of $D$ for water diffusion within the droplet, where the collective motion is substracted from molecular diffusion, is only ~$10^{-7}$cm$^2$/s at 200-400 K and ~$10^{-6}$ cm$^2$/s at 450 K (**Fig. 5c**), which is much lower than that in the bulk water at ambient condition ($D_0 = $~$10^{-5}$ cm$^2$/s) [2, 5]. This result confirms the solid-like behavior of EWs at the room temperature, and the occurrence of solid-fluid transition at ~480-490 K.

In MD simulations, the structures and thermodynamics of water predicted rely on the potential model used. Although there is no 'perfect' model that could capture all the complex behaviors of water, we verify our findings by performing comparative studies using the extended simple point charge model (SPC/E) model that is also widely used for nanoconfined water [7]. The results lead to the same conclusion in general. The water structures predicted are the same, although the transition temperature characterized using the SPC/E model, $T = $~400 K, is slightly lower.

## V. CONCLUSION

In brief, we explored the structure and thermodynamics encapsulated water covered by a graphene sheet. Our molecular dynamics simulation results show that the formation mechanism of layered structures is mainly attribute to the spatial confinement, where the graphene sheets offer an 'epitaxial' template, and the encapsulating pressure is critical for the appearance of in-plane orders. A first-order solid-fluid phase transition in 2D is identified for the encapsulated water, which is characterized by a discontinuous change in the entropy and structure of the hydrogen bond network, as well as an anomalous temperature dependence of the diffusion behavior. The wide temperature window of stable nanoconfined water structures under the cover of graphene sheet offers an ideal platform to explore the intriguing behaviors of water in a confined environment, and our findings here provide some fundamental understandings of it.

After the submission of current work, we notice a number of studies that report similar 2D ice structures of water confined between two parallel graphene sheets, which align with our type I and type II structures although the condition of confinement is different from our model [27-30]. The robustness of these 2D lattices highlights their significance for both theoretical interests and practical applications in nanoscale material and device design. Consequently, our structural and thermodynamics analysis of the EWs, as well as discussion on their molecular and collective diffusion add more understandings to these unique phases of water.

# ACKNOWLEDGMENTS

We thank Professors Shiang-Tai Lin and William A. Goddard III for the help on the free energy analysis using their 2PT program. This work was supported by the National Natural Science Foundation of China through Grants 11222217 and 11472150, and the State Key Laboratory of Mechanics and Control of Mechanical Structures (Nanjing University of Aeronautics and Astronautics) through Grant No. MCMS-0414G01. The computation was performed on the Explorer 100 cluster system at Tsinghua National Laboratory for Information Science and Technology.

## TABLES, FIGURES AND CAPTIONS

**Table 1.** Thermodynamic properties (entropy $S$, total energy $E$ and free energy $F$) at 300 K calculated for each water molecule under different confinements.

|  | $S$ (J/mol.K) | $E$ (KJ/mol) | $F$ (KJ/mol) |
|---|---|---|---|
| Partial interlayer intercalation[*] | 51.02 | -90.34 | -105.64 |
| Full interlayer intercalation[*] | 49.93 | -195.90 | -210.88 |
| Encapsulated water | 48.40 | -207.58 | -222.10 |

[*]For full intercalated water monolayer between graphene sheets, additional constraints come from the periodic boundary conditions applied in the lateral dimensions. Models are constructed as shown in **Fig. S3**, with an interlayer distance of 0.64 nm.

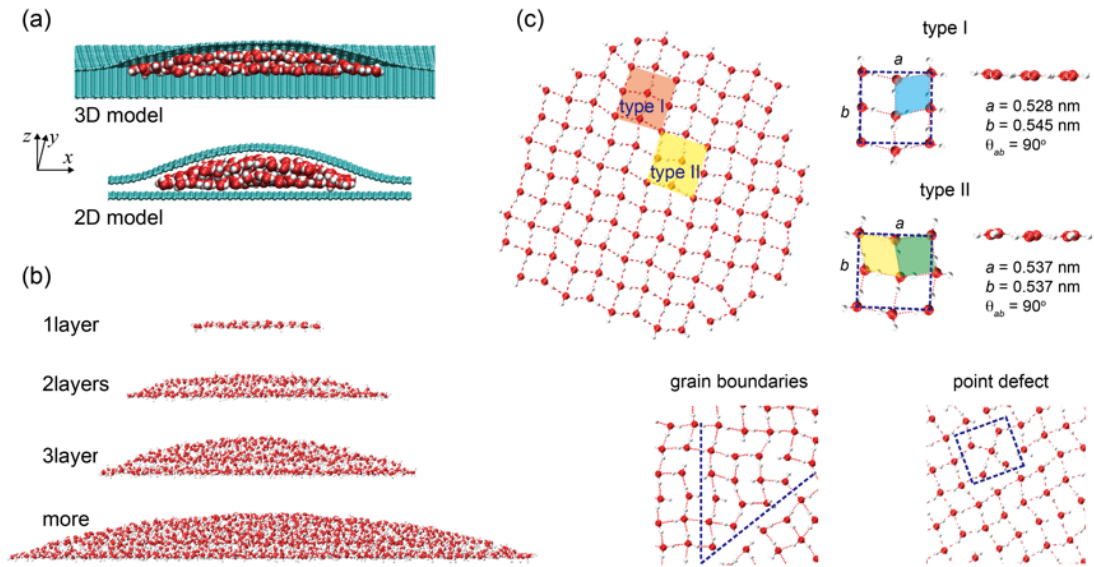

**Figure 1.** (a) Cross-section views of 3D and 2D models for water encapsulation under graphene. (b) Mono-, bi- and tri-layer water structures with the number of water molecules $N_W$ = 115, 546, 986 at 300 K, which disappear with $N_W$ > ~1000. (c) Simulation snapshots of monolayer water showing ordered structures of two different lattice types, as well as defects (grain boundaries and point defects). The lattice constants are denoted.

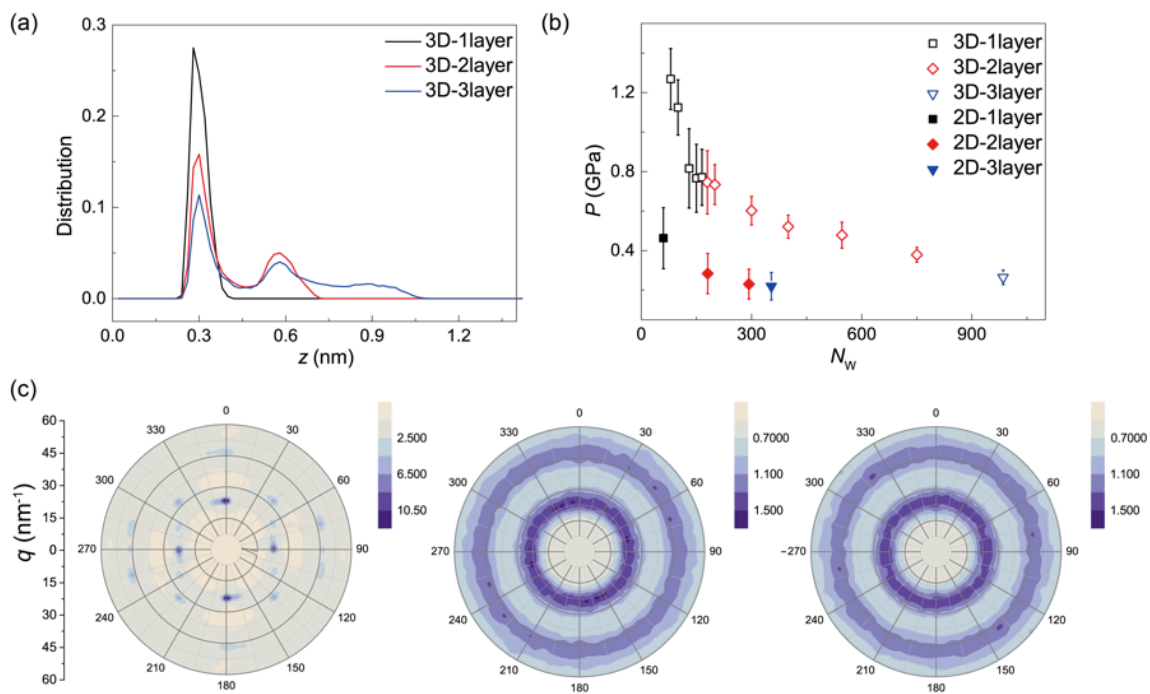

**Figure 2.** (a) Density profiles at 300 K, measured along the $z$ direction of encapsulated water in the 3D model. (b) Pressure in the water condensation plotted as a function of the number of water molecules at $T$ = 300 K. (c) Structural factors of nanoconfined mono-, bi- and tri-layer water with $N_{\mathrm{W}}$ = 115, 546, 986 at $T$ = 300 K. The error bars are obtained from the thermodynamic fluctuations measured in MD simulations. Here the wave vector **q** is represented as its amplitude $q$ and orientation in the polar plots. The color denotes the amplitude of structure factor $S(\mathbf{q})$.

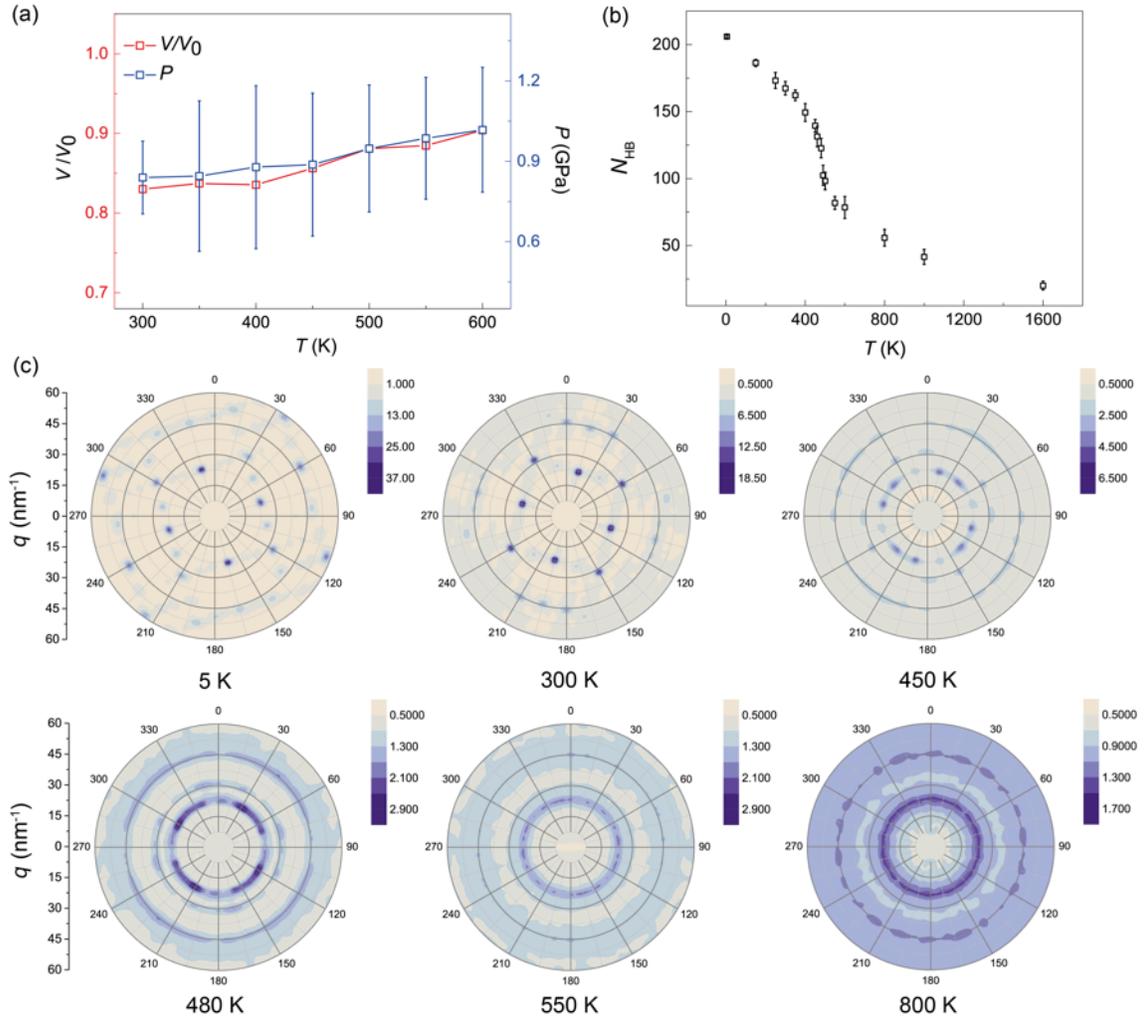

**Figure 3.** (a) Pressure, volume changes as function of $T$, where $V_0$ is volume of bulk water. (b) Number of H-bonds, $N_{HB}$, in the EW, plotted as a function of temperature. The error bars measure the thermodynamic fluctuations. (c) Structural factors of monolayer EW, demonstrating a solid-fluid phase transition at ~480-490 K.

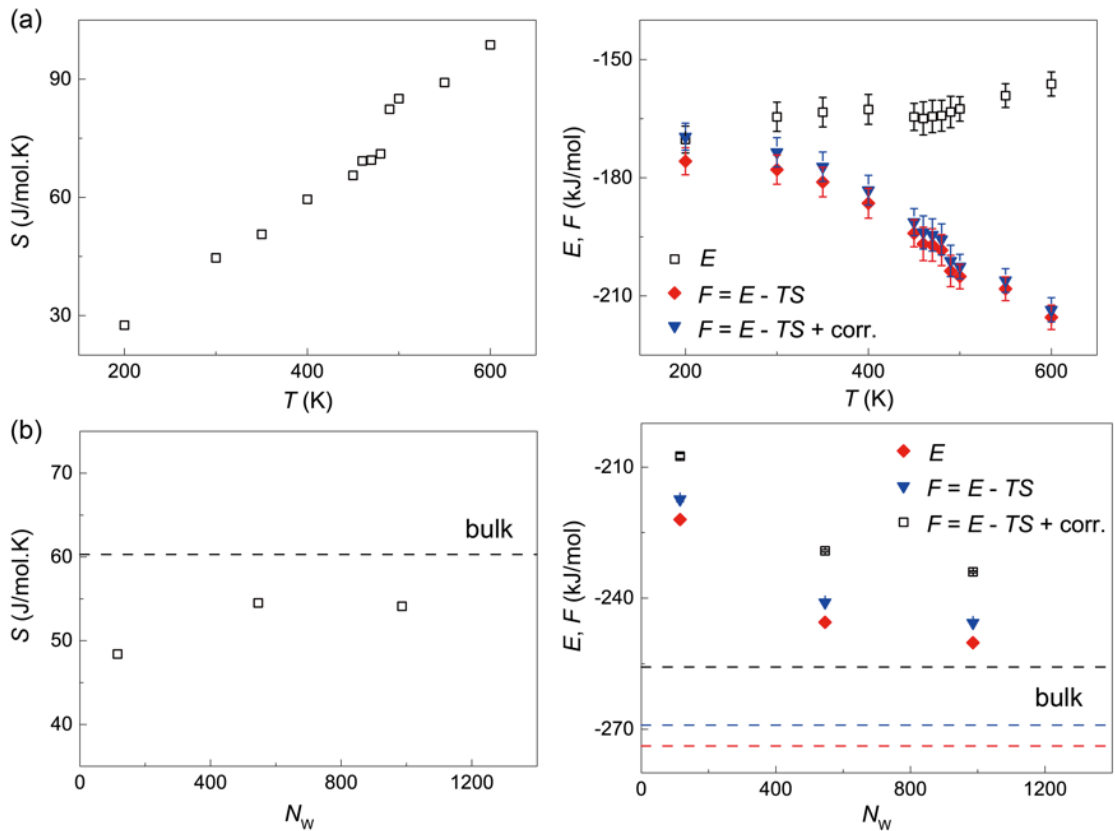

**Figure 4.** Entropy (*S*), total energy calculated from MD simulations (*E*), free energy (*F*) for the EW as temperature *T* = 200-600 K (a) and water number $N_W$ = 115, 546, 986 (mono-, bi-, tri-layer), as well as bulk water (b). The correction on free energy with ZPE and heat capacity terms are considered, using the 2PT model. The error bars measure the energy fluctuations in MD simulations.

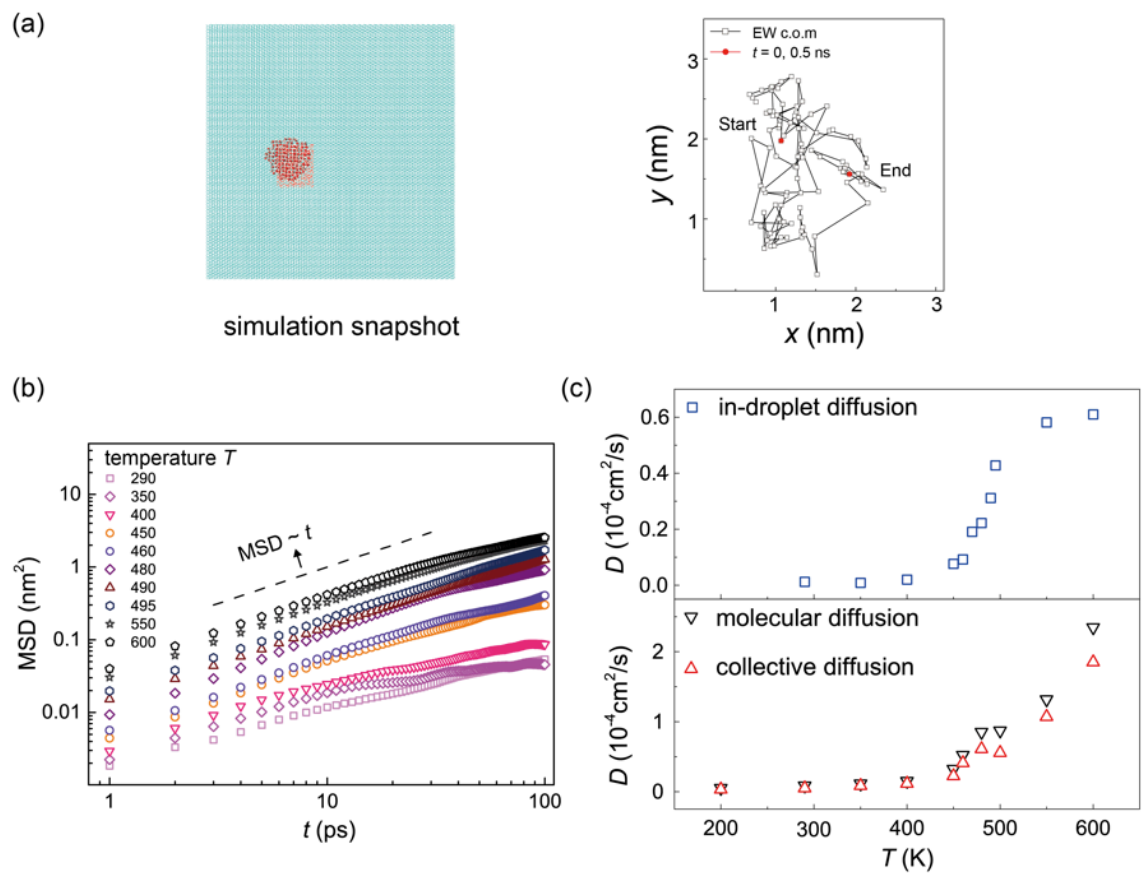

**Figure 5.** (a) A simulation snapshot and a 0.5 ns-long center-of-mass trajectory of the encapsulated water. Here only the graphene substrate is shown. (b) MSD of water molecular diffusion in a monolayer EW, calculated at different temperature. (c) Diffusion coefficient $D$ calculated for both molecular diffusion and the collective diffusion of EW. To calculate the in-droplet diffusion, the collective motion is substracted from molecular diffusion.

# REFERENCES

[1]     P. V. Hobbs, *Ice Physics* (Oxford University Press, 1974).

[2]     K. Koga, G. T. Gao, H. Tanaka, and X. C. Zeng, Nature 412, 802 (2001).

[3]     J. Hu, X.-D. Xiao, D. F. Ogletree, and M. Salmeron, Science 268, 267 (1995).

[4]     R. R. Nair, H. A. Wu, P. N. Jayaram, I. V. Grigorieva, and A. K. Geim, Science 335, 442 (2012).

[5]     S. Han, M. Y. Choi, P. Kumar, and H. E. Stanley, Nat. Phys. 6, 685 (2010).

[6]     J. Carrasco, A. Hodgson, and A. Michaelides, Nat. Mater. 11, 667 (2012).

[7]     G. Algara-Siller, O. Lehtinen, F. C. Wang, R. R. Nair, U. Kaiser, H. A. Wu, A. K. Geim, and I. V. Grigorieva, Nature 519, 443 (2015).

[8]     T. L. Hill, *Thermodynamics of Small Systems Part I & II* (Dover Publications, New York, 1994).

[9]     J. Shim, C. H. Lui, T. Y. Ko, Y.-J. Yu, P. Kim, T. F. Heinz, and S. Ryu, Nano Lett. 12, 648 (2012).

[10]    D. Lee, G. Ahn, and S. Ryu, J. Am. Chem. Soc. 136, 6634 (2014).

[11]    N. Severin, P. Lange, I. M. Sokolov, and J. P. Rabe, Nano Lett. 12, 774 (2012).

[12]    E. J. Olson, R. Ma, T. Sun, M. A. Ebrish, N. Haratipour, K. Min, N. R. Aluru, and S. J. Koester, ACS Appl. Mater. Interfaces 7, 25804 (2015).

[13]    S. Plimpton, J. Comp. Phys. 117, 1 (1995).

[14]    C.-J. Shih, S. Lin, R. Sharma, M. S. Strano, and D. Blankschtein, Langmuir 28, 235 (2011).

[15]    N. R. Tummala and A. Striolo, J. Phys. Chem. B 112, 1987 (2008).

[16]    N. Wei, C. Lv, and Z. Xu, Langmuir 30, 3572 (2014).

[17]    N. Samadashvili, B. Reischl, T. Hynninen, T. Ala-Nissilä, and A. S. Foster, Friction 1, 242 (2013).

[18]    N. Giovambattista, P. J. Rossky, and P. G. Debenedetti, Phys. Rev. Lett. 102, 050603 (2009).

[19]    C. Vega and J. L. F. Abascal, Phys. Chem. Chem. Phys. 13, 19663 (2011).

[20]    R. W. Hockney and J. W. Eastwood, *Computer Simulation Using Particles* (Taylor & Francis, 1989).

[21]    S. T. Lin, M. Blanco, and W. A. Goddard, J. Chem. Phys. 119, 11792 (2003).

[22]  K. Falk, F. Sedlmeier, L. Joly, R. R. Netz, and L. Bocquet, Nano Lett. 10, 4067 (2010).

[23]  W.-H. Zhao, L. Wang, J. Bai, L.-F. Yuan, J. Yang, and X. C. Zeng, Acc. Chem. Res. 47, 2505 (2014).

[24]  W.-H. Zhao, J. Bai, L.-F. Yuan, J. Yang, and X. C. Zeng, Chemical Science 5, 1757 (2014).

[25]  W. Zhu, T. Low, V. Perebeinos, A. A. Bol, Y. Zhu, H. Yan, J. Tersoff, and P. Avouris, Nano Lett. 12, 3431 (2012).

[26]  C. Lee, X. Wei, J. W. Kysar, and J. Hone, Science 321, 385 (2008).

[27]  Y. Zhu, F. Wang, J. Bai, X. C. Zeng, and H. Wu, ACS Nano  9, 12197 (2015).

[28]  J. Chen, G. Schusteritsch, C. J. Pickard, C. G. Salzmann, and A. Michaelides, Phys. Rev. Lett.  116, 025501 (2016).

[29]  F. Corsetti, J. Zubeltzu, and E. Artacho, Phys. Rev. Lett. 116, 085901 (2016).

[30]  F. Corsetti, P. Matthews, and E. Artacho, Sci. Rep. 6, 18651 (2016).